\UseRawInputEncoding

\def\PRL{{ Phys. Rev. Lett.\ }\/}
\def\PRB{{ Phys. Rev. B\ }\/}

\def\be{\begin {equation}}
\def\ee{\end {equation}}
\def\ber{\begin {eqnarray}}
\def\eer{\end {eqnarray}}
\def\bers{\begin {eqnarray*}}
\def\eers{\end {eqnarray*}}

\makeatletter

\newcommand{\Rmnum}[1]{\expandafter\@slowromancap\romannumeral #1@}
\makeatother

\makeatletter
\newcommand*\env@matrix[1][*\c@MaxMatrixCols c]{%
  \hskip -\arraycolsep
  \let\@ifnextchar\new@ifnextchar
  \array{#1}}
\makeatother

\documentclass[aps,prb,showpacs,superscriptaddress,a4paper,twocolumn]{revtex4-1}

\usepackage{amsmath}
\usepackage{float}
\usepackage{color}

\usepackage[normalem]{ulem}
\usepackage{soul}
\usepackage{amssymb}

\usepackage{amsfonts}
\usepackage{amssymb}
\usepackage{graphicx}
\begin {document}

\title{Co-existence of Topological Non-trivial and Spin Gapless Semiconducting Behavior in MnPO$_4$: A Composite Quantum Compound} 

\author{Chia-Hsiu Hsu}
\thanks{These two authors have contributed equally to this work}
\affiliation{Department of Physics, National Sun Yat-sen University, Kaohsiung 80424, Taiwan}

\author{P. C. Sreeparvathy}
\thanks{These two authors have contributed equally to this work}
\affiliation{Department of Physics, Indian Institute of Technology Bombay, Powai, Mumbai 400076, India}

\author{Chanchal K. Barman}
\affiliation{Department of Physics, Indian Institute of Technology Bombay, Powai, Mumbai 400076, India}

\author{Feng-Chuan Chuang}
\email{fchuang@mail.nsysu.edu.tw}
\affiliation{Department of Physics, National Sun Yat-sen University, Kaohsiung 80424, Taiwan}
\affiliation{Department of Physics, National Tsing Hua University, Hsinchu, 30043, Taiwan}
\affiliation{Physics Division, National Center for Theoretical Sciences, Hsinchu, 30013, Taiwan}

\author{Aftab Alam}
\email{aftab@iitb.ac.in}
\affiliation{Department of Physics, Indian Institute of Technology Bombay, Powai, Mumbai 400076, India}

\date{\today}

\begin{abstract}
Composite quantum compounds (CQC) are classic example of quantum materials which host more than one apparently distinct quantum phenomenon in physics. Magnetism, topological superconductivity, Rashba physics etc. are few such quantum phenomenon which are ubiquitously observed in several functional materials and can co-exist in CQCs. In this letter, we use {\it ab-initio} calculations to predict the co-existence of two incompatible phenomena, namely topologically non-trivial Weyl semimetal and spin gapless semiconducting (SGS)  behavior, in a single crystalline system. SGS belong to a special class of spintronics material which exhibit  a unique band structure involving a semiconducting state for one spin channel and a gapless state for the other. We report such a SGS  behavior in conjunction with the topologically non-trivial multi-Weyl Fermions in MnPO$_4$. Interestingly, these Weyl nodes are located very close to the Fermi level with the minimal trivial band density. A drumhead like surface state originating from a nodal loop around Y-point in the Brillouin zone is observed. A large value of the simulated anomalous Hall conductivity (1265 $\Omega^{-1} cm^{-1}$) indirectly reflects  the topological non-trivial behavior of this compound. Such co-existent quantum phenomena are not common in condensed matter systems and hence it opens up a fertile ground to explore and achieve newer functional materials. 
\end{abstract}

\maketitle


{\par}{\it \bf Introduction :}
Co-existence of two or more complimentary quantum phenomenon in a single material often provide a fertile ground to explore the fundamental correlation between these different processes in physics. Apart from basic science, such amalgamation of different quantum properties can also open the door for potential technological applications.  The compounds which show such combined quantum properties are called the composite quantum compounds (CQC).\cite{CQC} Till date, only a handful of such CQCs have been realized e.g. topological superconductivity involving  superconductivity and topology \cite{Hasan-Kane-RMP, Zhang-RMP} topological axion insulator between topology and magnetism, which are currently one of the emerging field of interest.\cite{axion-1, axion-2}

Spintronics and topological non-trivial band ordering are two distinct quantum phenomenon which can be connected via the common prerequisite of spin-orbit coupling (SOC). There are LiGaGe-type polar compounds where ferroelectricity and topological insulator (TI) phases are coupled together via the presence of strong SOC and broken inversion symmetry.\cite{rashba_weyl_dirac} Another classic example of CQC is BiTeI which shows co-existence of TI phase and large Rashba spin splitting under the influence of external pressure.\cite{BiTeI} Though there exist some reports on few CQCs, hunt for newer class with more exotic complimentary quantum properties is still ongoing. 

In this letter, we propose a material platform that can simultaneously showcase the existence of non-trivial multiple Weyl nodes and/or nodal-line along with the so-called spin gapless semiconducting (SGS) behavior.
SGS are new states of quantum matter which are characterised by a unique spin-polarized band structure.\cite{SGS} Unlike conventional semiconductors or half metallic ferromagnets, they carry a finite band gap for one spin channel and close (zero) gap for the other and thus are useful for tunable spin transport applications. It is one of the latest classes of materials considered for spintronic devices. A few of the several advantages of SGSs include (i) minimal amount of energy required to excite electrons from valence to conduction band due to zero gap (ii) availability of both charge carriers, i.e., electrons as well as holes, which can be 100\% spin-polarized simultaneously (iii) high Curie temperature for a large number of them.\cite{S_Skaftouros,Jiangchao_Han,Ashis_Kundu}
 Several number of SGS materials have been predicted in the last few years using both theoretical as well as experimental tools.\cite{SGS_our}
Weyl semimetals (WSM), on the other hand, are also known to show fascinating transport signatures e.g., anomalous Hall effect (AHE), which describes the transverse voltage drop resulting from an applied longitudinal electric field. Contrary to classical Hall effect under applied magnetic field, a momentum dependent fictitious magnetic field (Berry curvature) causes the AHE in Weyl semimetals.\cite{N$_$Nagaosa,R$_$Shindou}  As such, a magnetic Weyl semimetallic system could be an ideal platform to achieve a Berry curvature induced AHE because Weyl nodes are source(sink) of large Berry flux in this case. Another interesting feature is the large magnitude of anomalous Hall conductivity (AHC) in systems that host Weyl type band crossings near the Fermi level.\cite{W$_$Shi} For example, a large value of AHC  (=1130 $\Omega$$^{-1}$ cm$^{-1}$) is reported for Co$_{3}$Sn$_{2}$S$_{2}$.\cite{Enke$_$Liu,Jianlei$_$Shen} Co$_{2}$MnAl\cite{Peigang$_$Li} is another example which show a giant value of AHC, around 1277 $\Omega$$^{-1}$ cm$^{-1}$, at room temperature.  Several other magnetic compounds including Heusler alloys\cite{W$_$Shi, B$_$M$_$Ludbrook} are also found to show such Weyl points and consequently large AHC.\cite{Jianlei,Shubhankar,Guangqiang} The density functional theory calculations play a crucial role in discovering several topological semimetals. Yet, finding an ideal candidate material where Weyl/Dirac nodes lie close to the Fermi level (E$_F$) is always challenging.

The purpose of the present letter is to report MnPO$_4$ as the first CQC  which exhibit both SGS and multi-Weyl semimetal behavior. Unlike the conventional WSM, this compound shows four pairs of Weyl nodes in its band structure, all of which are closely located to E$_F$, with minimal trivial band density. It shows a significantly large AHC of 1265 $\Omega$$^{-1}$ cm$^{-1}$, similar to other reported WSM. Being the first of its kind, MnPO$_4$ is expected to enjoy all the advantages of the two unique quantum properties, as explained above. 
 


{\par}{\it \bf Computational Details :}
Vienna Ab-initio Simulation Package (VASP)\cite{G$_$Kresse1,G$_$Kresse2} based on density functional theory (DFT) have been used to carry out the simulation. We used generalized Gradient approximation by Perdew-Burke-Ernzerhof (PBE)\cite{G$_$Kresse2} to treat exchange and correlation. To enforce the localization of the Mn-d electrons, we perform PBE$+$U calculation\cite{hubardU} with a Hubbard U (=3.9 eV ) introduced in a screened Hartree-Fock manner. Justification for the use of this U value is given in the supplementary material(SI).\cite{supplement} Plane wave basis set using projector augmented wave (PAW)\cite{PEBloch} method was used with an energy cutoff of 520 eV. Brillouin zone (BZ) integrations were performed using a 12$\times$12$\times$10 k-mesh. Total energy was converged up to 10$^{-6}$ eV. {\bf The spin-orbit coupling (SOC)} effect was included. Phonon frequencies were calculated using Density Functional Perturbation Theory (DFPT) using phonopy.\cite{Togo} Maximally localized Wannier functions (MLWF)\cite{Nicola, Ivo$_$Souza, Nicola$_$Marzari} were use to construct a tight binding model to closely reproduce the bulk band structure.  WannierTools package\cite{QuanSheng} was employed to simulate the topological properties, and AHC.

\begin{table}[t]
	\caption{Relative energies ($\Delta$E) of MnPO$_{4}$ with different space groups (SG). Energy for Cmcm is taken as reference. }
	\begin{ruledtabular}
		\begin{tabular}{c c|c}
			Space groups  && $\Delta$E (meV/atom)\\
			\hline	
			Cmcm  && 0.0  \\
			P3$_{1}$21  &&70.0\\
			I$\overline{4}$ && 121.0  \\
		\end{tabular}
	\end{ruledtabular}
	\label{table:Genergy}
\end{table}
        
\begin{figure}[t]
\centering
	\includegraphics[width=\linewidth]{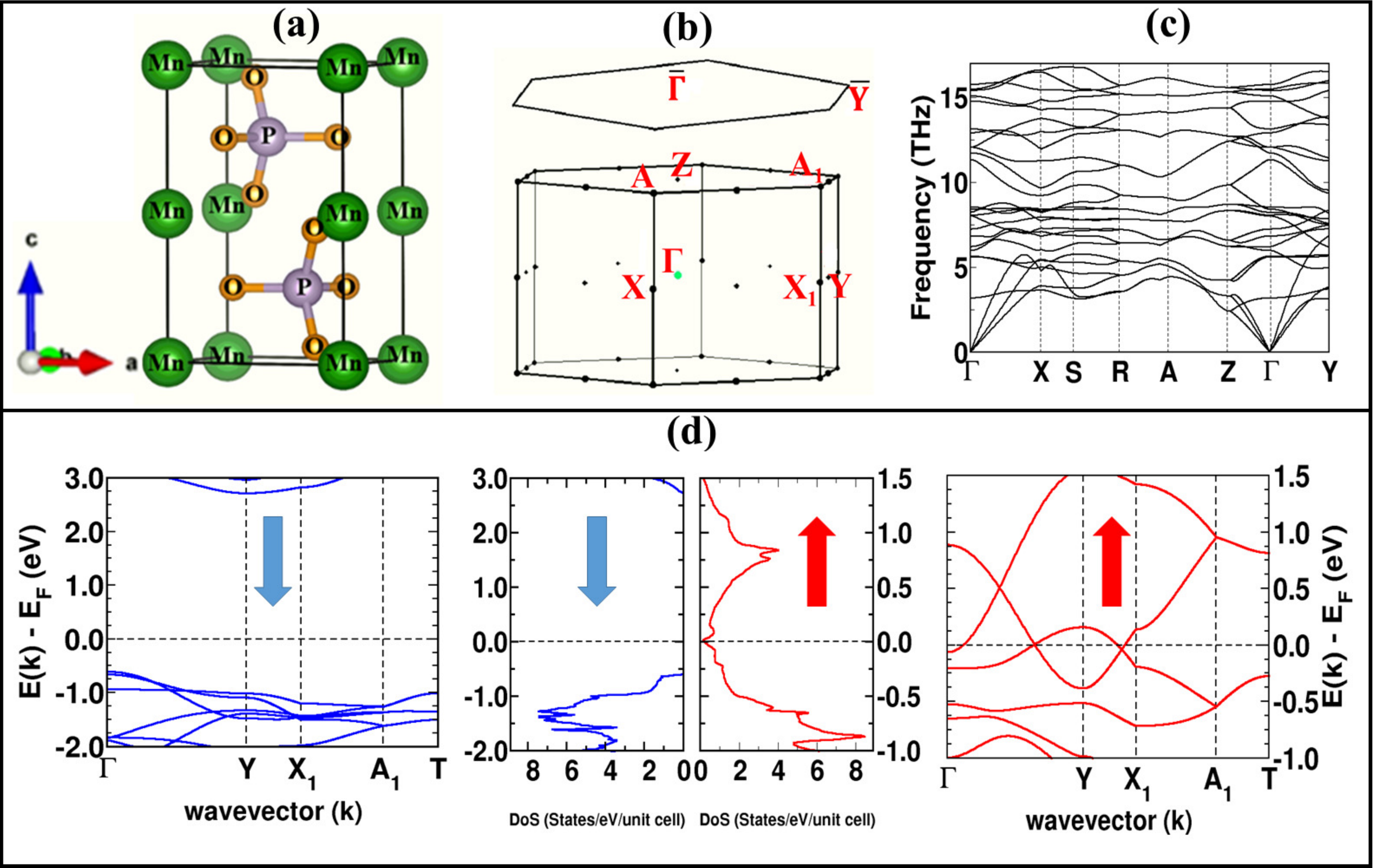}
\caption{(Color online) For MnPO$_{4}$ (a) Crystal structure with space group Cmcm (\# 63) (b) BZ for bulk and (001) surface (c) Phonon dispersion (d) Spin polarized band structure and density of states (DoS) without spin-orbit coupling (SOC).}
\label{fig1}
\end{figure}
        

\begin{center}
{\it \bf RESULTS AND DISCUSSION:}
\end{center}
{\it \bf  Crystal Structure and Stability:} MnPO$_{4}$ crystallizes in a structure with space group (SG) Cmcm (\# 63). One of the early experimental studies back in 1956 reported two additional sister structures of MnPO$_{4}$ (SG P3$_{1}$21 \& I$\overline{4}$).\cite{E$_$C$_$Shafer} However, Cmcm is energetically the most stable structure (see Table~\ref{table:Genergy}) and hence our choice for further calculations. The optimized lattice parameters of MnPO$_{4}$ in this structure (in primitive cell) is found to be $a=4.711$ \AA, $b=4.711$ \AA\ and $c=6.534$ \AA. The crystal structure itself is shown in Fig.~\ref{fig1}(a), which contains two formula units with a total of 12 atoms in the unit cell. Figure~\ref{fig1}(b) shows the corresponding hexagonal shaped BZ.
Dynamical stability of MnPO$_{4}$ (in this structure) has been verified by simulating the phonon dispersion, which does not show any imaginary phonon modes, as shown in Fig.~\ref{fig1}(c).

\begin{figure}[t]
	\centering
	\includegraphics[width=\linewidth]{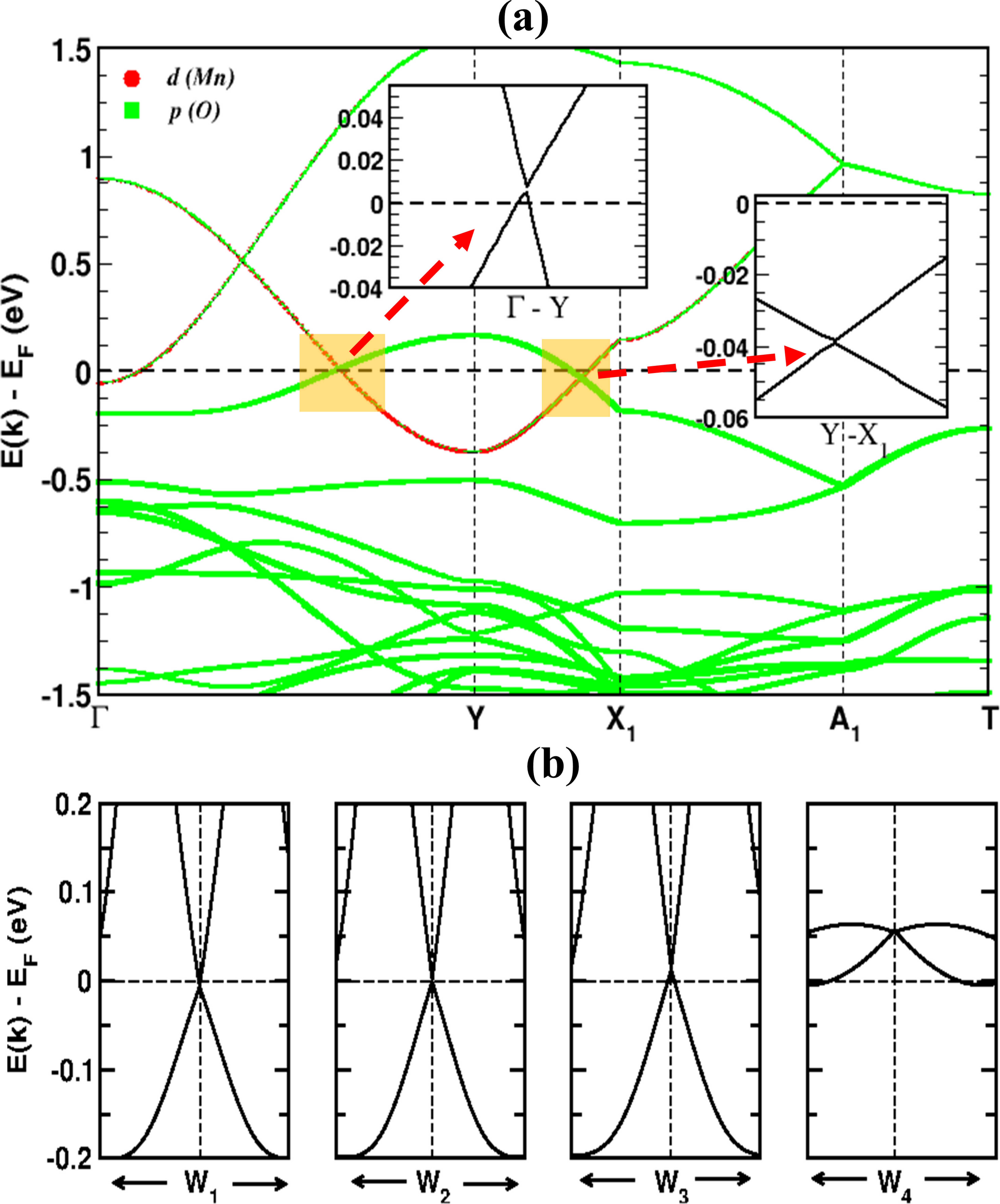}
	\caption{(a)Orbital projected band structure of MnPO$_4$ with SOC, indicating two topological nodal points, one along $\Gamma$-Y and the other along Y-X$_1$ high symmetry points. A zoomed view around the nodal points are shown in the insets. (b) Dispersion around four positive chiral Weyl nodes, W$_{i=1,2,3,4}^{+}$, with Chern number $+1$. The dispersion around Weyl node with negative Chern number looks very similar. }
	\label{figWeyl}
\end{figure}

{\it \bf  Electronic Structure:} 
 We simulated different magnetic ordering of Mn in MnPO$_{4}$, such as ferromagnetic (FM), Anti-ferromagnetic(AFM) and a few ferrimagnetic configurations using 2$\times$1$\times$1 supercell, which contains 4 Mn atoms (see SI\cite{supplement} for more details).  Among these, FM turns out be energetically the most stable one. The magnitude of exchange interactions between neighboring Mn atoms lie in the range $\sim$5-11 meV. We have also obtained a rough estimate of the Curie temperature, which turn out to be $\sim$505 K, implying a FM to paramagnetic  transition. Figure ~\ref{fig1}(d) shows the spin polarized band structure and density of states (without SOC) of MnPO$_4$ in the FM phase. It clearly confirms the SGS behavior where one spin channel is semiconducting while the other has a zero gap. The spin up band structure  shows an elliptical nodal ring type dispersion centering around the Y-point in BZ. However, there are tiny gap openings on the nodal ring along Y-$\Gamma$ direction. Figure~\ref{figWeyl}(a) shows the orbital projected band structure including the effect of SOC. The insets show a zoomed view at/near the nodal points. Inclusion of SOC does not change the band topology much but slightly enhances the energy gap along the nodal ring. Such small gap openings along the nodal ring have earlier been observed in many systems e.g., ZrSiS,\cite{ZrSiS2016-1,ZrSiS2016-2} Cu$_3$PdN,\cite{Cu3PdN-1,Cu3PdN-2} TiB$_2$,\cite{TiB2} CaPd,\cite{CaPd} CaAgBi,\cite{CaAgBi} etc. In the present case, the nodal ring is formed due to the band inversion between oxygen p-orbital and Mn-d orbital at/near Y-point.

\begin{table}[b]
	\caption{Weyl points location in the BZ and their respective chiralities. W$_{i}^{\pm}$ represent $i^{th}$ Weyl point with chirality $\pm1$.}
	\begin{ruledtabular}
		\begin{tabular}{|c|c|c|}
			
			Weyl points & Coordinates (k$_{x}$,k$_{y}$,k$_{z}$) & Chirality \\
			\hline	
			
			W$_{1}$$^{+}$  &(0.302,$-$0.404, 0.000)  & $+$1  \\
			W$_{1}$$^{-}$ &($-$0.302, 0.404, 0.000) & $-$1 \\
			W$_{2}$$^{+}$  &(0.381, $-$0.30, 0.000) & $+$1 \\
			W$_{2}$$^{-}$  &($-$0.385, 0.30, 0.000) & $-$1 \\
			W$_{3}$$^{+}$  &($-$0.334, 0.311, 0.000) & $+$1 \\
			W$_{3}$$^{-}$  &(0.338, $-$0.308, 0.000) & $-$1 \\
			W$_{4}$$^{+}$  &($-$0.499, 0.499, 0.233) & $+$1 \\
			W$_{4}$$^{-}$  &(0.499, $-$0.499, $-$0.233) & $-$1 \\
		\end{tabular}
	\end{ruledtabular}
	\label{table1}
\end{table}

Being ferromagnetic in nature, MnPO$_{4}$ lacks TRS and hence expected to contain Weyl like two-fold degenerate band crossings at a generic K-point in the BZ. Unlike other conventional Weyl semimetals, ferromagnetic MnPO$_{4}$ shows four pairs of Weyl nodes. Table~\ref{table1} shows the reciprocal space coordinates and the respective chiralities of these nodes. Interestingly, three-pairs of these nodes, (W$_i^{\pm}; i=1,2,3$) lie on $k_z=0$ plane, while the fourth one (W$_4^{\pm}$) on  $k_z=\pm0.233$ plane. The total topological charge associated with these eight Weyl nodes sum up to zero, which is in accordance with the Nielsen-Ninomiya theorem.\cite{NielsenNinomiya}  Eight Weyl points and nodal loop are marked in the BZ, and presented in suppelementary Figure S2.
Figure~\ref{figWeyl}(c) shows the  band dispersion near these four Weyl nodes.  

\begin{figure*}[t]
	\centering
	\includegraphics[width=0.9\linewidth]{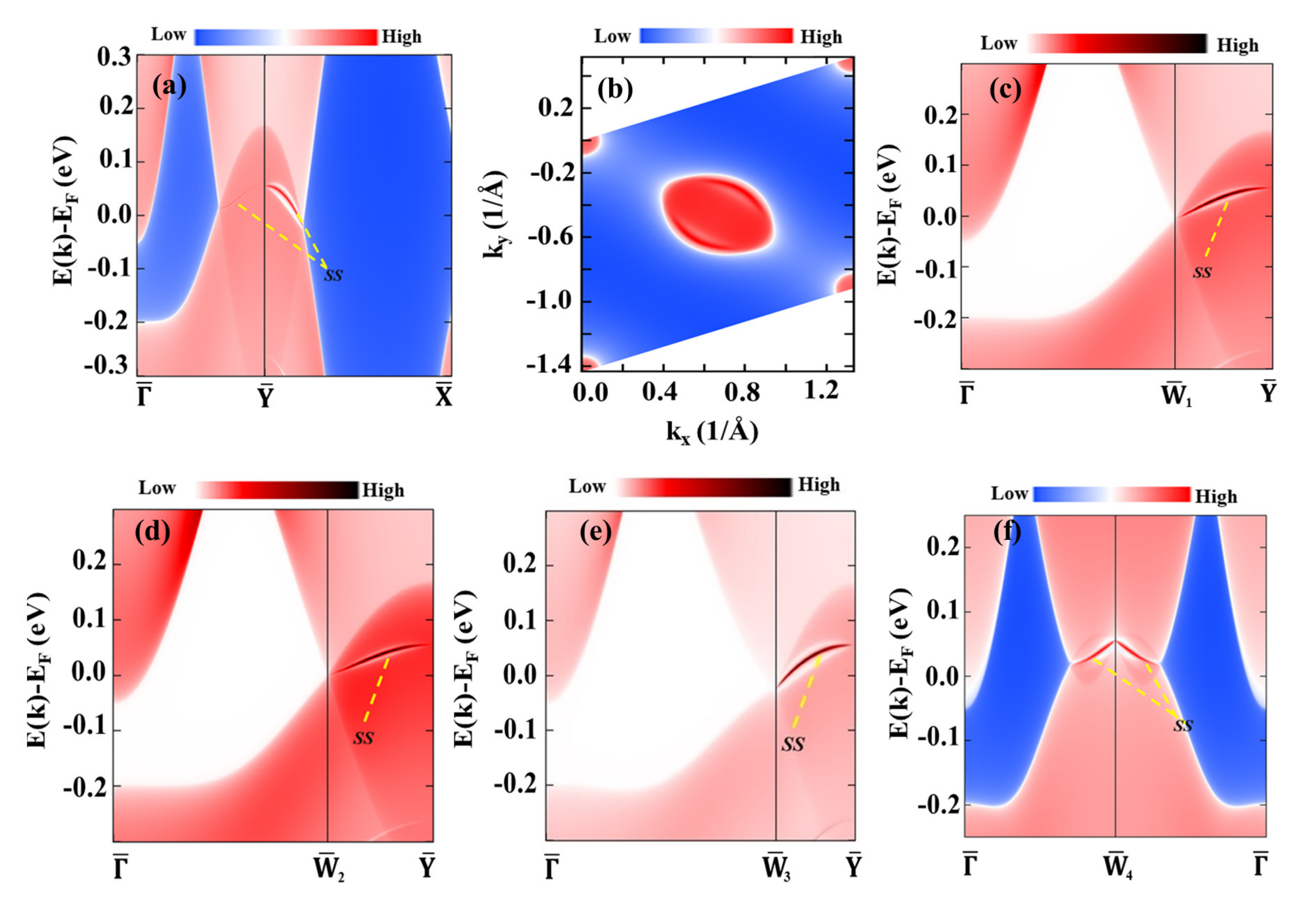}
	\caption{(Color online) Surface band structure of MnPO$_{4}$. (a) Electronic dispersion on the (001) surface around Y-point in the BZ.  SS indicates drumhead like surface states arising out of nodal ring, (b) Fermi surface topology on (001) surface. (c-e) Dispersion on (001) surface around three Weyl nodes, $\overline{W}_{i=1, 2, 3}$, (f) Dispersion on (010) surface of W$_{4}$.}
	\label{fig2}
\end{figure*}

{\par}{\it \bf  Surface States:} 
Surface states are one of the most enlightening features of a topological quantum material.  Based on the multi-faceted nature of the Weyl nodes in the bulk, MnPO$_{4}$ is expected to host rich surface states. We have analyzed the projected (001)-surface band structures. Due to the nodal line type behavior of the bulk electronic structure, a ``{\it drumhead}'' like unique surface state appears when the nodal loop is projected on the (001) surface. The corresponding surface states and Fermi surface topology are presented in Fig.~\ref{fig2}(a,b). As evident from Fig.~\ref{fig2}(a), bright surface states (drumhead like) are clearly distinguishable from the bulk counter part. Also Fermi surface at an iso-energy of E$_{F}$ display the elliptical nodal ring  on the (001) surface as expected (see Fig.~\ref{fig2}(b)). Further, since nodal ring breaks down under the influence of SOC with a subsequent  appearance of Weyl nodes, one should also expect the emergence of surface states from the Weyl points. 
The surface dispersions near W$_{i=1, 2, 3}$ Weyl nodes are simulated on the (001) surface, while that for W$_4$, it is performed on the (010) surface. This is becaue the projcetion of W$_4$ on (001) surface aligned very close to the Y point. These surface dispersions are shown in Fig.~\ref{fig2}(c-f). One can observe the bright surface states emerging from each of these nodes as well. The Fermi arcs arising out of these Weyl points are, however, not seen on the (001) surface since they are submerged within the bulk spectral density.

{\par}{\it \bf  Anomalous Hall Effect:} 
A magnetic Weyl semimetal is often characterized by it's transverse conductivity. Here, we have examined the AHC of MnPO$_4$ in a wide range of energy span.
Since the magnetization of the compound is oriented  along the \emph{z-}direction, the principal transport parameter  to analyze here is  $\sigma_{xy}$. We have computed the anomalous Hall conductivity from Berry curvature, using the following relation, 

\begin{eqnarray*}
\sigma_{\alpha\beta} = -\frac{e^2}{\hbar}\int_{BZ}\frac{d^3k}{(2\pi)^3}\Omega_{\alpha\beta}(k)
\end{eqnarray*}

\begin{figure}[b]
	\centering
	\includegraphics[width=1.0\linewidth]{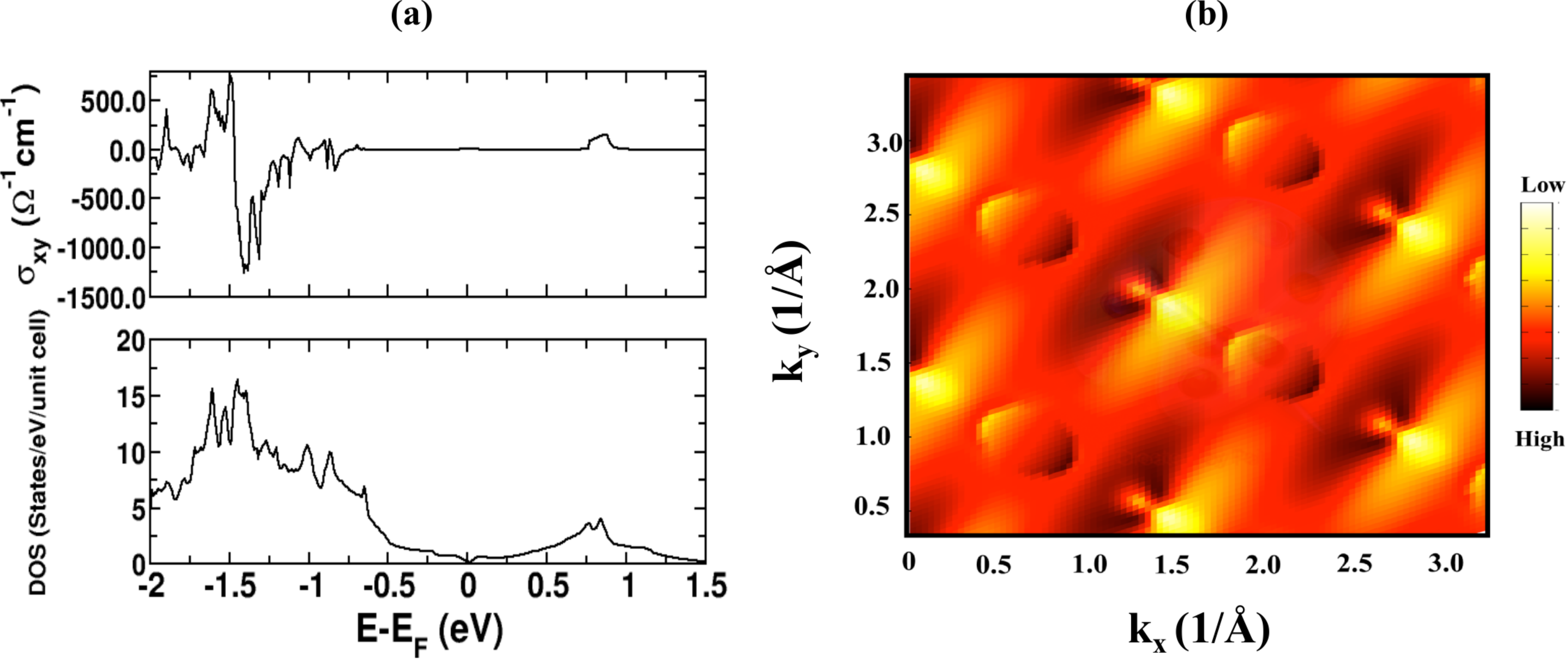}
	\caption{ (Color online) (a) Anomalous Hall conductivity (top), density of states (bottom), including SOC contribution. (b) z-component of Berry curvature at E$_F$ on the k$_{x}$-k$_{y}$ plane. }
	\label{fig4}
\end{figure}

Figure \ref{fig4}(a) shows the energy dependence of AHC for ferromagnetic MnPO$_{4}$. There are few important energy range where  the variation/magnitude of AHC is quite interesting and hence require a discussion. Between -0.5 to +0.5 eV, $\sigma_{xy}$ is almost negligible, which increases to 100 $\Omega$$^{-1}$ cm$^{-1}$ in the range -1.0 to -0.6 eV, and further to 220 $\Omega$$^{-1}$ cm$^{-1}$ at around -0.8 eV. These later values of $\sigma_{xy}$ are comparable with the corresponding reported values in several Heusler alloys.\cite{Husmann, Binoy, Kudrnovsky} Most promisingly, we have observed a peak in  $\sigma_{xy}$ at around -1.37 eV with the largest magnitude of 1265 $\Omega$$^{-1}$ cm$^{-1}$. Generally, such peaks in AHC is expected to occur at an energy ( E$_{F}$ in the case of MnPO$_{4}$) where Weyl points occur. However, in the present case, it is occurring somewhat away from the E$_{F}$. Such behavior is reported in few other materials as well in the past.\cite{Charles, Kaustuv2} To better understand this feature of AHC, a close inspection of electronic structure is essential. One plausible reason for such behavior is the lack of large density of states near E$_{F}$ (see Fig.~ \ref{fig4}(a) lower panel), which can eventually provide a decremental effect on both Berry curvature and AHC. \cite{EI} Apart from this, we further plotted the z-component of Berry curvature in Fig. \ref{fig4}(b). Here, light yellow color represents positive hot spot, and dark brown represents the negative hot spot,  and from the  Fig. \ref{fig4}(b), one can clearly see the lower magnitude of Berry curvature.
The percentage of spin polarization also influences the net magnitude of AHC in weak SOC systems. \cite{Kaustuv, TJen} The total AHC ($\sigma_{xy}^{t}$) can be written as the sum of $\sigma_{xy}$ (spin up)  and  $\sigma_{xy}$ (spin down) at specific chemical potential. \cite{Kaustuv, TJen} For our proposed compound MnPO$_{4}$, a 100 $\%$ spin polarization is observed at E$_{F}$, which remains preserved for a wide energy range of -0.6  to +2.7 eV. This actually suppresses the AHC values. Below -0.6 eV, one can clearly see the mixture of spin up and down states, causing an enhanced value of AHC. Similar trend has been reported in few Co-based Heusler alloys.\cite{Charles, Kaustuv2} It is clearly evident from the above discussion that MnPO$_{4}$ is indeed a promising candidate for future spintronic application. It can even become promising for topological  materials related applications if one can tune the position of Fermi level by doping or external pressure/strain without much compromising the band profile.

{\par}{\it \bf  Summary:} 
In summary, we report the co-existence of two complimentary quantum phenomenon, namely spin gapless semiconducting(SGS) behavior and topologically non-trivial multi-Weyl Fermions, in a single compound MnPO$_4$. Materials with such co-existent phenomenon are called composite quantum compounds (CQC). SGSs are a special class of spintronic materials which acquire a unique band structure with one spin channel showing semiconducting nature while the other shows a gapless state (simlar to graphene). Our detailed band structure calculations unfold the existence of SGS as well as \emph{four} distinct pairs of Weyl nodes in the bulk electronic dispersion of MnPO$_4$.  Interestingly, these Weyl nodes are located very close to the Fermi level with the minimal trivial band density. The chirality of these Weyl nodes are confirmed by calculating their respective topological charges. We found the so called ``drumhead'' like nodal line surface states on the (001) surface.
 Such co-existent topological semimetallic signature along with the SGS behavior in MnPO$_4$, will definitely attract further research interests in the field of spintronics and topological non-trivial materials.

{\par}{\it \bf  Acknowledgements:} 
AA acknowledges DST SERB, India (Grant No. CRG/2019/002050) for funding to support this research.  SPC and CKB  thank IIT Bombay spacetime computing facility, and Dr.Debashish Das  for helpful discussions. F.C.C acknowledges support from National Center for Theoretical Sciences and the Ministry of Science and Technology of Taiwan under Grants No. MOST-107-2628-M-110-001-MY3. He is also grateful to the National Center for High-performance Computing for computer time and facilities.




\begin{thebibliography}{113}

\bibitem{CQC} J. Li, Y. Li, S. Du, Z. Wang, B-L. Gu, S-C. Zhang, K. He, W. Duan, and Y. Xu, Science Advances {\bf 5}, eaaw5685 (2019).

\bibitem{Hasan-Kane-RMP} M. Z. Hasan, and C. L. Kane, Rev. Mod. Phys. {\bf 82}, 3045-3067 (2010).

\bibitem{Zhang-RMP} X. L. Qi, and S. C. Zhang, Rev. Mod. Phys. {\bf 83}, 1057-1110 (2011).

\bibitem{axion-1} C. Liu, Y. Wang, H. Li, Y. Wu, Y. Li, J. Li, K. He, Y. Xu, J. Zhang, and Y. Wang,  Nat. Mater.  {\bf 19}, 522-527 (2020).

\bibitem{axion-2} Y. S. Hou, J. W. Kim, and R. Q. Wu,  \PRB  {\bf 101}, 121401(R) (2020).
\bibitem{rashba_weyl_dirac} D. D. Sante, P. Barone, A. Stroppa, K. F. Garrity, D. Vanderbilt, and S. Picozzi, \PRL {\bf 117}, 076401 (2016). 

\bibitem{BiTeI} X. Xi, C. Ma, Z. Liu, Z. Chen, W. Ku, H. Berger, C. Martin, D. B. Tanner, and G. L. Carr, Phys. Rev. Lett. {\bf 111}, 155701 (2013); M.S. Bahramy, B.J. Yang, R. Arita, and N. Nagaosa, Nat. Commun. {\bf 3}, 679 (2012).

\bibitem{SGS} S. Quardi, G. H. Fecher, C. Felser and J. Kubler, Phys. Rev. Lett. {\bf 110}, 100401 (2013); X. L. Wang, Phys. Rev. Lett. {\bf 100}, 156404 (2013).

\bibitem{S_Skaftouros} S. Skaftouros,  K. Özdoğan,  E. Şaşıoğlu and  I. Galanakis, Appl. Phys. Lett. {\bf 102}, 022402 (2013).

\bibitem{Jiangchao_Han} Jiangchao Han, Yulin Feng, Kailun Yao and G. Y. Gao, Appl. Phys. Lett. {\bf111}, 132402 (2017).
\bibitem{Ashis_Kundu}Ashis Kundu, Srikrishna Ghosh, Rudra Banerjee, Subhradip Ghosh, Biplab Sanyal, Sci. Rep. {\bf7}, 1803. (2017).

\bibitem{SGS_our} D. Rani, Enamullah, L. Bainsla, K. G. Suresh and A. Alam, Phys. Rev. B {\bf 99}, 104429 (2019); D. Rani, L. Bainsla, A. Alam and K. G. Suresh, J. Appl. Phys. 128, 220902 (2020); L. Bainsla, A. I. Mallick, M. Manivel Raja, A. K. Nigam, B. S. D. Ch. S. Varaprasad, Y. K. Takahashi, Aftab Alam, K. G. Suresh, and K. Hono, Phys. Rev. B {\bf 91}, 104408 (2015);  L. Bainsla, A. I. Mallick, M. Manivel Raja, A. A. Coelho, A. K. Nigam, D. D. Johnson, Aftab Alam,  and K. G. Suresh, Phys. Rev. B {\bf 92}, 045201 (2015).  














\bibitem{N$_$Nagaosa} N. Nagaosa, J. Sinova, S. Onoda, A. H. MacDonald, and N. P. Ong, Rev. Mod. Phys. {\bf 82}, 1539 (2010).


\bibitem{R$_$Shindou}  R.  Shindou  and  N.  Nagaosa, Phys. Rev. Lett. {\bf87},  116801 (2001).


\bibitem{W$_$Shi} W. Shi, L. Muechler, K. Manna, Y. Zhang, K. Koepernik, R. Car, J. van den Brink, C. Felser, and Y. Sun, Phys. Rev. B {\bf97},  060406 (2018).


\bibitem{Enke$_$Liu}Enke Liu, Yan Sun, Nitesh Kumar, Lukas Muechler, Aili Sun, Lin Jiao, Shuo-Ying Yang, Defa Liu, Aiji Liang, Qiunan Xu, Johannes Kroder, Vicky Süß, Horst Borrmann, Chandra Shekhar, Zhaosheng Wang, Chuanying Xi, Wenhong Wang, Walter Schnelle, Steffen Wirth, Yulin Chen, Sebastian T. B. Goennenwein, Claudia Felser, Nat. Phys. {\bf14}, 1125–1131 (2018). 


\bibitem{Jianlei$_$Shen} Jianlei Shen, Qingqi Zeng, Shen Zhang, Hongyi Sun, Qiushi Yao, Xuekui Xi,  Wenhong Wang, Guangheng Wu, Baogen Shen, Qihang Liu, and Enke Liu, Adv. Funct. Mater., {\bf30}, 2000830 (2020).

\bibitem{Peigang$_$Li}Peigang Li, Jahyun Koo, Wei Ning, Jinguo Li, Leixin Miao, Lujin Min, Yanglin Zhu, Yu Wang, Nasim Alem, Chao-Xing Liu, Zhiqiang Mao, Binghai Yan, Nat. Commun. {\bf11}, 3476 (2020).



\bibitem{B$_$M$_$Ludbrook} B. M. Ludbrook, B. J. Ruck, and S. Granville, Appl. Phys. Lett. {\bf 110}, 062408 (2017).


\bibitem{Jianlei} Jianlei Shen, Qiushi Yao , Qingqi Zeng , Hongyi Sun, Xuekui Xi , Guangheng Wu, Wenhong Wang
Baogen Shen,  Qihang Liu and Enke Liu, Phys, Rev. Lett. {\bf125}, 086602 (2020).

\bibitem{Shubhankar}Shubhankar Roy, Ratnadwip Singha, Arup Ghosh, Arnab Pariari and Prabhat Mandal, Phys. Rev. B {\bf102}, 085147 (2020).

\bibitem{Guangqiang} Guangqiang Wang, Zhanghao Sun, Xinyu Si, and Shuang Jia, Chin. Phys. B {\bf29}, 077503 (2020).













\bibitem{G$_$Kresse1}G. Kresse and J. Hafner, Phys. Rev. B {\bf47}, 558(R) (1993).
\bibitem{G$_$Kresse2}G. Kresse and D. Joubert, Phys. Rev. B {\bf59}, 1758 (1999).

\bibitem{hubardU} S.  L. Dudarev, G.  A. Botton, S.  Y. Savrasov, C.  J. Humphreys, and A.  P. Sutton, Phys. Rev. B {\bf 57}, 1505 (1998).

\bibitem{PEBloch} P. E. Bl\"{o}chl, Phys. Rev. B {\bf 50}, 17953 (1994).

\bibitem{Togo}Atsushi Togo and Isao Tanaka, Scripta Materialia {\bf108}, 1-5 (2015).

\bibitem{Nicola} Nicola Marzari and David Vanderbilt, Phys. Rev. B {\bf56}, 12847 (1997).

\bibitem{Ivo$_$Souza}Ivo Souza, Nicola Marzari, and David Vanderbilt, Phys. Rev. B {\bf65}, 035109 (2001).

\bibitem{Nicola$_$Marzari}Nicola Marzari, Arash A. Mostofi, Jonathan R. Yates, IvoSouza, and David Vanderbilt, Rev. Mod. Phys. {\bf84}, 1419 (2012).
\bibitem{QuanSheng} QuanSheng Wu and ShengNan Zhang and Hai-Feng Song and Matthias Troyer and Alexey A. Soluyanov, Computer Physics Communications, {\bf224}, 405 - 416 (2018).
\bibitem{E$_$C$_$Shafer} E. C. Shafer, M. W. Shafer and Rustum Roy, Zeitschrift für Kristallographie, Bd. {\bf 107}, 263—275 (1956).

\bibitem{supplement} See Supplementary Material at [URL] for more details.


\bibitem{ZrSiS2016-1} Schoop, L. M,  Mazhar N. Ali, Carola Straßer,Andreas Topp,  Andrei Varykhalov, Dmitry Marchenko, Viola Duppel, Stuart S. P. Parkin, Bettina V. Lotsch and Christian R. Ast, Nat. Commun. {\bf 7}, 11696 (2016).


\bibitem{ZrSiS2016-2}Neupane, M. Ilya Belopolski, M. Mofazzel Hosen, Daniel S. Sanchez, Raman Sankar,Maria Szlawska, Su-Yang Xu, Klauss Dimitri, Nagendra Dhakal, Pablo Maldonado, Peter M. Oppeneer, Dariusz Kaczorowski, Fangcheng Chou, M. Zahid Hasan, and Tomasz Durakiewicz, Phys. Rev. B {\bf 93}, 201104 (2016).


\bibitem{Cu3PdN-1}Kim, Y., Wieder, B. J., Kane, C. L. \& Rappe, Phys. Rev. Lett. {\bf 115}, 036806 (2015).

\bibitem{Cu3PdN-2} Yu, R. Hongming Weng, Zhong Fang, Xi Dai and Xiao Hu, Phys. Rev. Lett. {\bf 115}, 036807 (2015).

\bibitem{TiB2} Liu, Z., Rui Lou, Pengjie Guo, Qi Wang, Shanshan Sun, Chenghe Li, Setti Thirupathaiah, Alexander Fedorov, Dawei Shen, Kai Liu, Hechang Lei, and Shancai Wang, Phys. Rev. X {\bf 8}, 031044 (2018).


\bibitem{CaPd}Liu, G., Jin, L., Dai, X., Chen, G. \& Zhang, X., Phys. Rev. B {\bf 98}, 075157 (2018).

\bibitem{CaAgBi}Yamakage, A., Yamakawa, Y., Tanaka, Y. \& Okamoto, Y., J. Phys. Soc. Jpn. {\bf 85}, 013708 (2016).

\bibitem{NielsenNinomiya} H. B. Nielsen and M. Ninomiya, Phys. Lett. B {\bf 105}, 219 (1981).

\bibitem{Husmann} A. Husmann and L. J. Singh, Phys. Rev. B {\bf 73}, 172417 (2006). 

\bibitem{Binoy}Binoy Krishna Hazra, M. Manivel Raja, R. Rawat, Archana Lakhani, S.N. Kaul, S. Srinath, Journal of Magnetism and Magnetic Materials {\bf448}, 371–377  (2018).

\bibitem{Kudrnovsky} J. Kudrnovsk´y and V. Drchal, I. Turek, Phys. Rev. B {\bf88}, 014422 (2013). 

\bibitem{Charles} Charles S. Spencer, Jacob Gayles, Nicholas A. Porter, Satoshi Sugimoto, Zabeada Aslam, Christian J. Kinane,
Timothy R. Charlton, Frank Freimuth, Stanislav Chadov, Sean Langridge, Jairo Sinova, Claudia Felser,
Stefan Blügel, Yuriy Mokrousov, and Christopher H. Marrows, Phys. Rev. B {\bf97}, 214406 (2018). 
\bibitem{Kaustuv2} Kaustuv Manna, Yan Sun, Lukas Muechler, Jürgen Kübler and Claudia Felser, Nat. Rev. Mater., {\bf3}, 244–256 (2018). 
\bibitem{EI} E. I. Kondorskii, Zh. Eksp. Teor. Fiz. {\bf 55}, 558 (1968).

\bibitem{Kaustuv}Kaustuv Manna, Lukas Muechler, Ting-Hui Kao, Rolf Stinshoff, Yang Zhang, Johannes Gooth, Nitesh Kumar,
Guido Kreiner, Klaus Koepernik, Roberto Car, Jürgen Kübler, Gerhard H. Fecher, Chandra Shekhar,
Yan Sun, and Claudia Felser, Phys. Rev. X {\bf 8}, 041045 (2018).
\bibitem{TJen}T. Jen-Chuan and G. Guang-Yu, New J. Phys. {\bf15}, 033014 (2013).


\end{thebibliography}
\end{document}


\title{Supplementary Material \\  Co-existence of Topological Non-trivial and Spin Gapless Semiconducting Behavior in MnPO$_4$: A Composite Quantum Compound } 

\author{Chia-Hsiu Hsu}
\thanks{These two authors have contributed equally to this work}
\affiliation{Department of Physics, National Sun Yat-sen University, Kaohsiung 80424, Taiwan}

\author{P. C. Sreeparvathy}
\thanks{These two authors have contributed equally to this work}
\affiliation{Department of Physics, Indian Institute of Technology Bombay, Powai, Mumbai 400076, India}

\author{Chanchal K. Barman}
\affiliation{Department of Physics, Indian Institute of Technology Bombay, Powai, Mumbai 400076, India}

\author{Feng-Chuan Chuang}
\email{fchuang@mail.nsysu.edu.tw}
\affiliation{Department of Physics, National Sun Yat-sen University, Kaohsiung 80424, Taiwan}
\affiliation{Department of Physics, National Tsing Hua University, Hsinchu, 30043, Taiwan}
\affiliation{Physics Division, National Center for Theoretical Sciences, Hsinchu, 30013, Taiwan}

\author{Aftab Alam}
\email{aftab@iitb.ac.in}
\affiliation{Department of Physics, Indian Institute of Technology Bombay, Powai, Mumbai 400076, India}

\date{\today}

\begin{abstract}
 
\end{abstract}

\maketitle

\subsection{Computational methods}
To confirm the Hubbard U value, we have checked the dependence of several properties (e.g. magnetic moment, position of topological nodal points etc.) with varying U values. The net magnetic moment of the cell changes within 2$\%$ (for U values varying between 0 to 6 eV). On the other hand, the position of topological nodal points (along $\Gamma$-Y $\&$ Y-X1, see Fig. 2(a) of manuscript) varies within 0.06 meV as a function of U. We chose U=3.9 eV because this U-value locates the nodal points as close as possible to the Fermi level. We have used Dudarev's approach\cite{dudarev}, where it is U$_{eff}$ =(U-J) which goes into the calculation and hence J was not varies individually.

\begin{figure}[b!]
	\centering
	\includegraphics[width=0.9\linewidth]{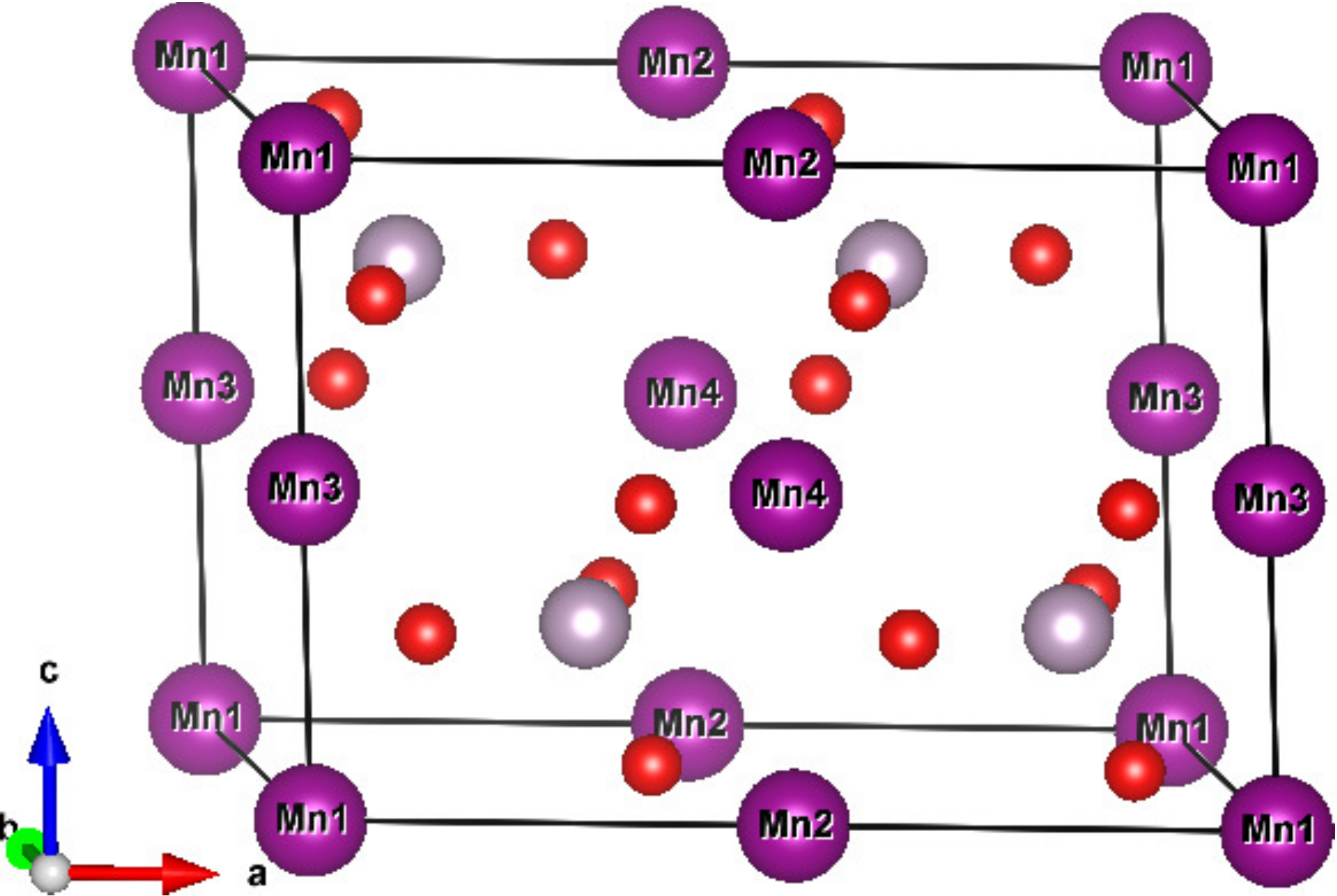}
	\caption{(a) 2$\times$1$\times$1  supercell structure of MnPO$_{4}$}
	\label{structure}
\end{figure}

\begin{table}[b!]
	\centering
	\small\addtolength{\tabcolsep}{-1pt}
	\caption{Energy difference($\Delta$E in eV) and magnetic moment for different magnetic ordering}
	\begin{ruledtabular}
		\begin{tabular}{|c|c|c|}
			
			Phase &$\Delta$E &Moment(moment/atom($\mu$$_{B}$)) \\
			\hline   &     &                                   \\
			FM     &0.00  &3.99                                  \\
			AFM    &+0.089 & 0.00                                \\
			Ferrimagnetic  &+0.041 &1.99                          \\
			\hline  &   &                                          \\			
		\end{tabular}
	\end{ruledtabular}
	\label{table1}
\end{table}

\subsection{Magnetic phase}
To check the energetically most favorable magnetic ordering for Mn in MnPO$_{4}$, we have considered a 2$\times$1$\times$1 supercell, which contains 4 Mn atoms. This supercell structure is shown in Fig.~\ref{structure} below. The relative energies of FM, AFM and a Ferrimagnetic phase is tabulated in Table~\ref{table1}.  AFM ordering corresponds to spin up at Mn$_{1}$ and Mn$_{2}$ while spin down at Mn$_{3}$ and Mn$_{4}$ sites.  Ferrimagnetic ordering corresponds to spin up at Mn$_{1}$, Mn$_{3}$ and Mn$_{4}$   and spin down at Mn$_{2}$ site. Although, we have simulated few other magnetic configurations as well (not shown here), the ones which are most competing are sown in Table~\ref{table1}. Clearly, FM is energetically the most stable ordering, and hence chosen for the rest of the studies.
To estimate the magnetic exchange interactions between Mn atoms, we have considered three pairs i.e. Mn$_{1}$-Mn$_{3}$ (nearest neighbor), Mn$_{1}$-Mn$_{2}$ and Mn$_{3}$-Mn$_{4}$ (next nearest neighbor). 
The calculated exchange parameters between these pairs are J$_{12}$ = 0.0051 eV, J$_{13}$ = 0.0112 eV and J$_{34}$ = 0.0051 eV. The corresponding Curie temperature is ~505.2 K, which can mediate the possibility of FM to paramagnetic transition in MnPO$_{4}$.  

\begin{figure}[hb!]
	\centering
	\includegraphics[width=0.4\linewidth]{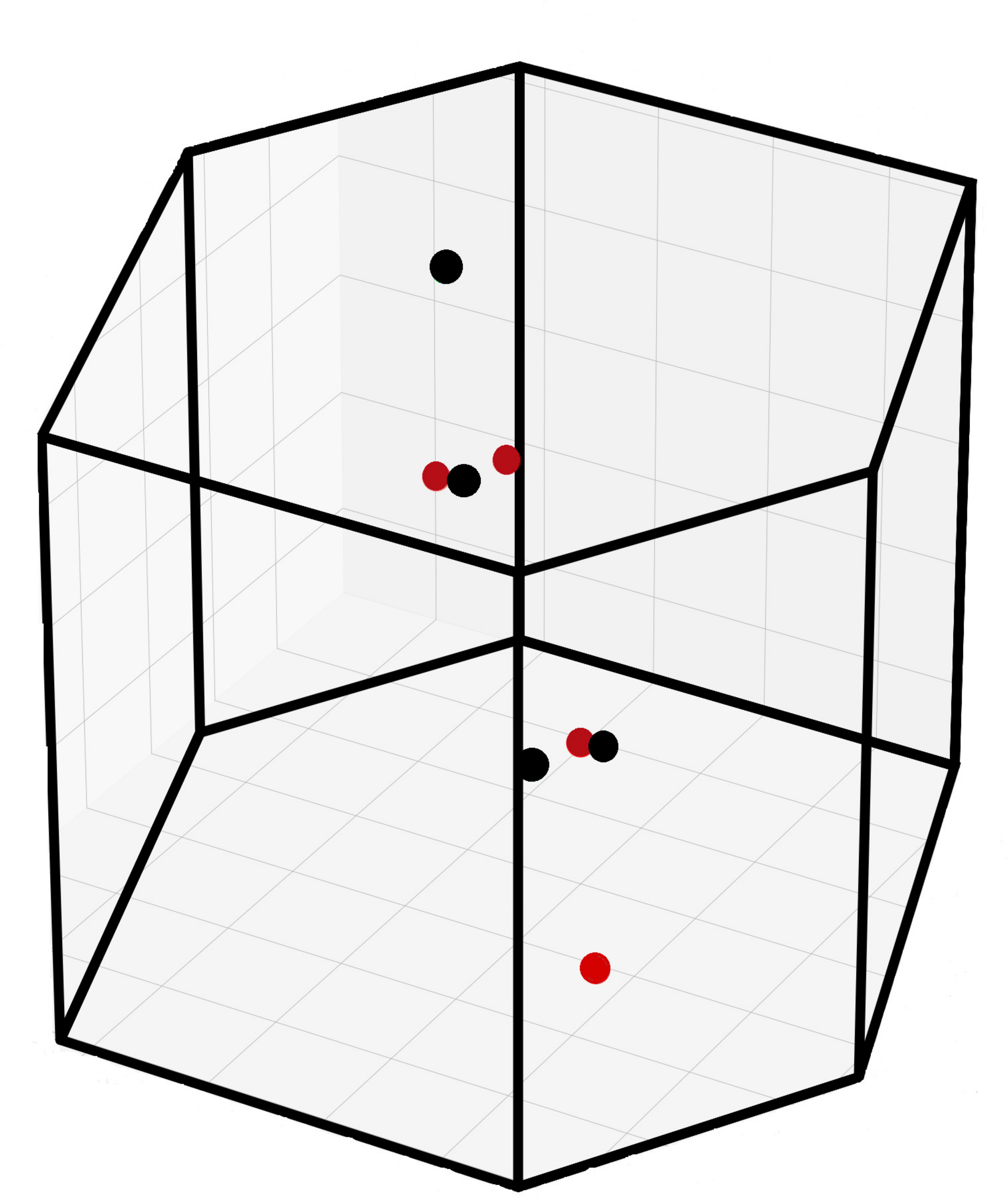}
	\includegraphics[width=0.4\linewidth]{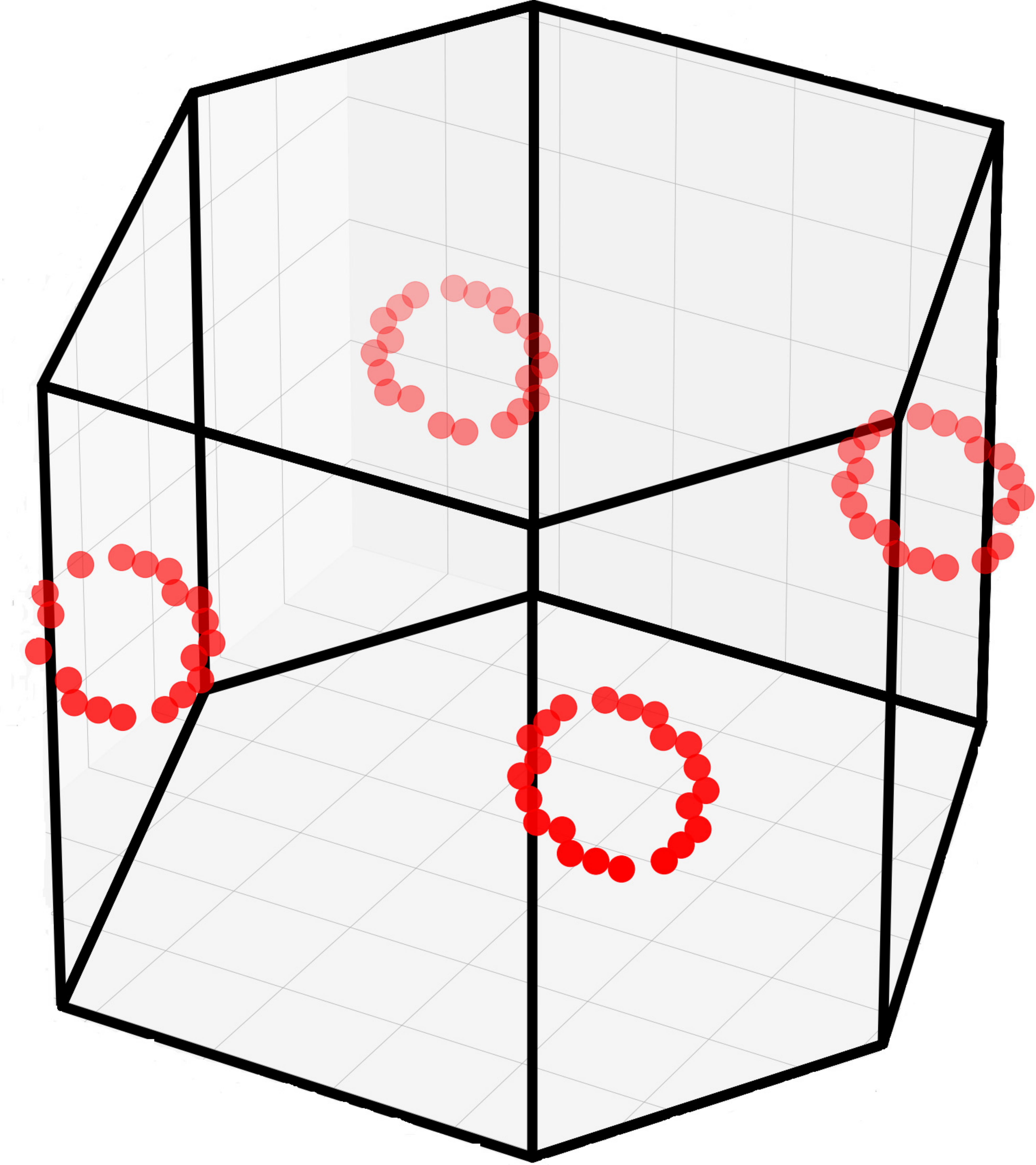}
	\caption{Locus of eight Weyl nodes (left) and elliptical nodal ring (right) in the bulk BZ.}
	\label{figweyl}
\end{figure}